\documentclass[twocolumn,showpacs,preprintnumbers,amsmath,amssymb]{revtex4}

\usepackage{bm}
\usepackage{color,graphicx}

\begin{document}

\title{Relativistic Spin and Dirac Spin in relativistically covariant Stern-Gerlach Experiment}

\author{Taeseung Choi}
\address{Division of General Education, Seoul Women's University, Seoul 139-774 }
\address{School of Computational Sciences, Korea Institute for Advanced Study, Seoul 130-012, Korea }
\email{tschoi@swu.ac.kr}
\author{Yeong Deok Han}
\address{Department of Game Contents, Woosuk University, Wanju, Cheonbuk, 565-701, Republic of Korea}
\email{ydhan@woosuk.ac.kr}


\begin{abstract}
     We have studied a relativistically covariant Stern-Gerlach (SG) experiment for a relativistic spin and a Dirac spin. 
We have obtained the relativistic spin in an arbitrary frame by using the classical spin dipole tensor, 
which gives the covariant spin dipole interactions, and the relation between a spin and a spin magnetic dipole moment. 
The relativistic spin is shown to have problems to become a proper spin operator for a massive relativistic particle 
because of two reasons. First, the relativistic spin three-vector operators cannot satisfy the spin algebra. 
Second, the SG experiment for the relativistic particle provides a paradox between two observers in the particle rest frame 
and the laboratory frame, in which the particle is moving. 
We have shown that the paradox in the SG experiment is resolved by the Dirac spin, which is covariantly defined 
by a Lorentz transformation in the Dirac spinor representation. 
The Dirac spin three-vector operators satisfy the spin algebra. 
It is shown that the SG experiment for the Dirac spin in the inertial frame, where there is only magnetic field, can determine 
the spin without the information of the momentum of the particle. 
     This shows that the reduced spin density matrix for the Dirac particle can be well-defined by integrating out 
     the momentum degrees of freedom.
\end{abstract}


\maketitle

\section{Introduction}

Recently high research intensities have emerged from the field known as relativistic quantum information
\cite{Czachor, Peres, Alsing, Gingrich, Li, Ahn, Czachor1, Terashima, Peres2, Lee, Kim, 
Caban1, Jordan, Lumata, Jordan2, Landulfo, Dunningham,Caban2, Friis, Choi,
Fuentes, Alsing1, Fuentes1, Palmer}. 
One of the typical examples to study relativistic effects on quantum information is the spin of 
a massive Dirac particle (e.g., electron). 
Spin and momentum degrees of freedom of a massive Dirac particle are entangled in general. 
Moreover, a Lorentz transformation can change the entanglement between them. 
This fact has created a controversy in understanding the spin under a Lorentz transformation. 
The problem of defining reduced spin density matrix for a massive Dirac particle, which encodes the 
quantum information of the spin, by integrating out the momentum degree of freedom is still controversial 
\cite{Peres, Gingrich, Li,Peres2, Lee,Caban1, Jordan, Lumata, Jordan2, Landulfo, Dunningham, Friis, Saldanha}. 

At this stage we need a clear understanding of the spin operator itself in an arbitrary reference frame. 
Even though the reduced spin density matrix must be a superposition of eigenstates of the spin operator, 
tracing over the momentum degree of freedom, which is entangled with the spin degree of freedom, 
makes the clear understanding of the spin itself difficult. 
Therefore, mathematical formalisms related with the reduced density matrix are not enough to clarify 
the controversial arguments in previous works. 
To understand the spin operator physically, the investigation of the spin measurement is most desirable. 
The best candidate is the famous Stern-Gerlach (SG) experiment, in which a spin interacts with 
electromagnetic fields. In the original SG experiment, the interaction is considered as non-relativistic. 
Special relativity requires that every observer must observe the same experimental result, 
which is the clicking event of the detector. 
We call this requirement a (relativistic) covariance of the SG experiment.

Recently Saldanha et al. presented an interesting new treatment for the spin of a massive relativistic particle 
in the covariant generalization of the Stern-Gerlach (SG) experiment \cite{Saldanha}. 
They have tried to detour the problem of defining the spin in an arbitrary reference frame by using the 
covariance of the SG Hamiltonian. 
The discussion of the SG experiment in the particle rest frame seems to be clear, 
because the spin operator in the particle rest frame is well-known as the usual Pauli spin operator. 
The covariance of the spin dipole interaction in the SG experiment, however, inevitably requires the discussion 
of the relativistic spin in an arbitrary frame. 

Usually, the spin has been considered as a relativistic object because the two-dimensional 
spinor representation is an irreducible representation of the Lorentz group. Also, a spin is 
the second Casimir of the inhomogeneous Lorentz (Poincar$\acute{\mbox{e}}$) group so that the unitary representation 
of the spin for the Poincar$\acute{\mbox{e}}$ group is the Wigner's little group.  
Hence the relativistically covariant generalization of the SG Hamiltonian is expected to give the proper 
definition of the spin operator in an arbitrary frame. 
The manifestly covariant form of dipole interactions can be obtained by the dipole tensor defined in the classical electrodynamics \cite{Penfield}. 
By using the Lorentz transformation of the magnetic dipole moment one can obtain the spin in an arbitrary frame. 
We will call this spin the relativistic spin. 
The definition of the relativistic spin in an arbitrary frame can be extracted from the spin magnetic dipole moment in that 
frame through the relation between a spin magnetic dipole moment and a spin for a free relativistic particle \cite{Zawadzki}. 


The critical issue, however, is that the origin of the spin for a massive Dirac particle 
is not only relativistic but also quantum mechanical. As is well-known, the spin index of a massive 
Dirac particle is naturally introduced by the Dirac theory that has successfully reconciled quantum 
theory with special relativity \cite{Ryder1}. 
This suggests that the proper spin operator for a massive Dirac particle must be defined by using the 
Dirac theory. In the Dirac theory, the spinor representation is not 2-dimensional but 4-dimensional. This means that 
the Dirac spinor is not an irreducible representation of the Lorentz group itself, 
but an irreducible representation of the Lorentz group extended by parity \cite{Ryder1}. 
Therefore, the symmetry group for a massive Dirac particle is not just the Poincar$\acute{\mbox{e}}$ group 
but must be extended by parity. 
This means that the spin operator for a massive Dirac particle in an arbitrary frame must be obtained 
in the 4-dimensional Dirac spinor representation. 
We will call this spin operator the Dirac spin operator to distinguish from the relativistic spin operator. 

We will show in this paper that the relativistic spin obtained by the classical dipole tensor 
cannot be a spin operator because the relativistic spin three-vector operators do not satisfy the spin algebra. 
Furthermore, the covariance of the SG experiment provides a paradox for the relativistic spin among 
observers in different motions. 
To resolve the paradox it is essential to use the Dirac spin operator. 
The Dirac spin operator will be shown to have additional pure quantum mechanical terms to the relativistic 
spin operator.
These pure quantum mechanical terms are critical to satisfy the spin algebra and to interpret 
the SG experiment covariantly and consistently.


\section{Relativistic spin and Dirac spin}
\label{sec:CSGH}

In this section we will study the relativistic spin and the Dirac spin. 
The spin tensors for both spins are well-defined on Minkowski space-time. 
The relativistic spin tensor in an arbitrary frame will be obtained by the relativistic generalization of spin dipole 
interactions in classical electrodynamics. On the other hand, the Dirac spin tensor in an arbitrary frame will be 
obtained by a Lorentz transformation in the Dirac spinor representation from the spin tensor in the particle rest frame. 

\subsection{Relativistic spin}
\label{subsec:RS}

The Hamiltonian of the original SG experiment is given by non-relativistic classical magnetic dipole interaction. 
In this sense we will call the relativistic generalization of the non-relativistic SG interaction, classical. 
A covariance of the SG experiment is required by special relativity, because all observers in their own inertial frames 
must observe the same events for the detectors of the SG devices. 
In relativistic case we must consider both electric and magnetic interactions with spin dipole moments, 
because a magnetic field in one reference frame would transform to electric and magnetic fields in another reference frame.

The covariant treatment of the SG Hamiltonian in classical electrodynamics is given by the antisymmetric 
dipole moment tensor \cite{Penfield}:
\begin{eqnarray}
{D}_{\alpha\kappa} = \gamma \left( \begin{array}{cccc} 0 &  {d}^1 & {d}^2 &  {d}^3 \\
- {d}^1 & 0 &  {\mu}^3 & - {\mu}^2 \\ - {d}^2 & - {\mu}^3 & 0 &  {\mu}^1 \\
- {d}^3 &  {\mu}^2 & - {\mu}^1 & 0 \end{array}\right),
\end{eqnarray} 
which is the same in Ref. \cite{Saldanha}. 
Greek indices $\alpha$ and $\kappa$ run from $0$ to $3$.
We define the electromagnetic field tensor as 
\begin{eqnarray}
{F}^{\alpha\kappa}=
\left( \begin{array}{cccc} 0 & - {E}^1 & -{E}^2 & -{E}^3 \\
{E}^1 & 0 & -{B}^3 & {B}^2 \\ {E}^2 & {B}^3 & 0 & - {B}^1 \\
{E}^3 & -{B}^2 & {B}^1 & 0 \end{array}\right),
\end{eqnarray}
where ${B}^i$ and ${E}^i$ are $i$-components of the magnetic and the electric fields, respectively.

The classical SG Hamiltonian in an arbitrary frame can be written with the manifestly covariant form 
\begin{eqnarray}
\label{Eq:CRSG}
{\mathcal{H}}_{\scriptsize{\mbox{SG}}} =- \frac{1}{2\gamma} F^{\alpha\kappa} D_{\alpha\kappa},
\end{eqnarray}
where $\gamma=1/\sqrt{1- {\boldsymbol{\beta}}^2}$ is the Lorentz factor and ${\boldsymbol{\beta}}={\bf v}/c$. 
We use the natural unit $\hbar=c=1$ and the metric tensor $g^{\alpha\kappa}=\mbox{diag}(+,-,-,-)$. 
Then $\boldsymbol{\beta}$ can be represented by the velocity ${\bf v}$ of the Lorentz boost. 

The field tensor and the dipole moment tensor transform under Lorentz transformation according to 
\begin{eqnarray}
\label{Eq:DTT}
F^{\alpha \kappa} = L^\alpha_{\phantom{\rho}\rho}L^\kappa_{\phantom{\delta}\lambda} \tilde{F}^{\rho \lambda} 
\mbox{ and }
D_{\alpha \kappa} = L_\alpha^{\phantom{\rho}\rho}L_\kappa^{\phantom{\delta}\lambda} \tilde{D}_{\rho \lambda},
\end{eqnarray}
where a tilde symbol $~\tilde{}~$ will be used to represent quantities in the particle rest frame. 
$L^\alpha_{\phantom{\mu}\rho}$ is the pure boost that relates the 
four-momentum $p^\alpha=(\gamma m, \gamma m v^x,\gamma m v^y, \gamma m v^z)$ in the moving frame to 
the four-momentum $\tilde{p}^\alpha=(m,0,0,0)$ in the particle rest frame by 
$p^\alpha = L^\alpha_{\phantom{\mu}\rho}\tilde{p}^\rho$. 
It is enough to consider the pure boost since the rotation is trivial for the spin. 

The electric dipole moment ${\bf d}$ and the magnetic dipole moment ${\boldsymbol{\mu}}$ in a moving frame 
are obtained from those in the particle rest frame as 
\begin{eqnarray}
\label{Eq:RTRDM}
{\bf d} &=&  \tilde{\bf d}+    (\boldsymbol{\beta}\times {\tilde{\boldsymbol{\mu}}})
- \frac{\gamma \boldsymbol{\beta} (\boldsymbol{\beta}\cdot{\tilde{\bf{d}}})}{\gamma+1} \\ \nonumber
{\boldsymbol{\mu}} &=&   \tilde{\boldsymbol{\mu}}-   (\boldsymbol{\beta}\times \tilde{\bf{d}})
- \frac{\gamma \boldsymbol{\beta} (\boldsymbol{\beta}\cdot\tilde{\boldsymbol{\mu}})}{\gamma+1},
\end{eqnarray} 
by using the Lorentz transformation and the relations 
\begin{eqnarray}
\label{Eq:DVT}
d^i = \frac{D_{0i}}{\gamma}, ~~~~~~ \mu^i =  \frac{1}{2 \gamma} \epsilon_{ijk}D_{jk},
\end{eqnarray}
where $\epsilon_{ijk}$ is the Levi-civita symbol. 
In this paper we denote a contravariant three-vector by bold-face letter such as ${\bf A}=(A^1, A^2, A^3)$.

Then the relativistic SG Hamiltonian in Eq. (\ref{Eq:CRSG}) can be rewritten as the usual form in 
classical electrodynamics: 
\begin{eqnarray}
\label{Eq:CSGH}
{\mathcal{H}}_{\scriptsize{\mbox{SG}}} = - \boldsymbol{{\mu}} \cdot \bf{{B}} -{\bf{{ d}}}\cdot {\bf{{E}}}.
\end{eqnarray}    
We assume that the particle has no electric and magnetic charges but only a spin to focus on the properties of the spin. 
In this case the classical intrinsic electric dipole moment, which is the classical electric dipole moment in the particle 
rest frame, is zero.
Then only a spin magnetic dipole interaction occurs in the particle rest frame.

Now we will define the relativistic spin in an arbitrary frame by using the relation between the spin and the spin 
magnetic dipole moment for a massive relativistic particle \cite{Zawadzki}. 
The relation is $\boldsymbol{\mu}=\alpha {\bf S}$ in the non-relativistic limit, where ${\bf S}$ is a spin 
three-vector and $\alpha$ is the gyromagnetic ratio with 
the rest mass of the particle in the denominator. 
In the relativistic generalization to an arbitrary moving frame, the rest mass is replaced by the relativistic mass 
such that the magnetic dipole moment $\boldsymbol{\mu}$ will be $\alpha {\bf S}/\gamma$. 
To give this relation, the relativistic spin tensor in a moving frame must be defined as
\begin{eqnarray}
\label{Eq:CRST}
{S}_{\mu\nu} = \frac{1}{\alpha} D_{\mu\nu}.
\end{eqnarray} 
Therefore, the relativistic spin three-vector operator is obtained as 
\begin{eqnarray}
\label{Eq:CRS}
{\bf S} = \gamma \tilde{\bf S} -\frac{\gamma^2 \boldsymbol{\beta} (\boldsymbol{\beta}\cdot\tilde{\bf{S}})}{\gamma+1},
\end{eqnarray} 
with the definition of 
\begin{eqnarray}
\label{Eq:SVT}
S^i = \frac{1}{2}\epsilon^{ijk} S_{jk}.
\end{eqnarray} 
Roman indices $i$, $j$, and $k$ runs from $1$ to $3$. 

Notice that in the particle rest frame the representation of the spin operator $\tilde{S}^i$ is ${\sigma}^i/2$, where 
${\sigma}^i$ is a Pauli matrix. 
One can check that the relativistic spin three-vector operators ${\bf S}$ do not satisfy the spin algebra 
$[S^i, S^j]= \epsilon^{ijk}S^k$. 
This means that the relativistic spin operator cannot be a proper spin operator.



\subsection{Dirac spin}

The massive spin-1/2 particle is well-known to be described by the Dirac equation, which has successfully reconciled 
quantum theory with special relativity. In this sense the massive relativistic particle with spin-1/2 is called the 
massive Dirac particle. 
We will call the spin operator for the massive Dirac particle the Dirac spin operator. 
The spin index is naturally introduced in the Dirac spinor representation \cite{Ryder1}. 
This means that the origin of the Dirac spin is not only relativistic but also quantum mechanical.  

In this paper we will use the covariant formalism of the Dirac theory. 
The covariant Dirac equation for a massive Dirac particle in momentum representation is 
\begin{eqnarray}
\label{Eq:DE}
(p^\mu \gamma_\mu -m)\psi=0,
\end{eqnarray}
where $\gamma^\mu$ are Dirac matrices:  
\begin{eqnarray}
\gamma^0 = \left(
\begin{array}{cc} I & 0 \\
0 & -I \end{array} \right), ~~
\gamma^i=  \left( \begin{array}{cc} 0 &  \sigma^i \\ -\sigma^i & 0 \end{array} \right),
\end{eqnarray}
in the standard representation. 
In this paper we will work in the standard spinor representation, because we are interested in 
the one-particle theory, where there is no superposition of positive- and negative-energy states. 
A Dirac spinor, which is a solution of the Dirac equation in Eq. (\ref{Eq:DE}) can be denoted 
as $|p,\sigma\rangle $ \cite{Ryder1}. 
The $p$ and $\sigma$ are momentum and spin indices, respectively. 

To determine the spin operator, we first consider the spin tensor operator $\mathbb{S}_{\alpha\kappa}$, 
which satisfies the commutation relations of the angular momentum tensor. 
The spin tensor operator transforms under the Lorentz boost $L$ as 
\begin{eqnarray}
\label{Eq:STTR}
\mathcal{D}(L) \mathbb{S}_{\alpha\kappa} \mathcal{D}^{-1}(L)= L^{\rho}_{\phantom{\mu}\alpha} 
L^\lambda_{\phantom{\mu}\kappa} \mathbb{S}_{\rho\lambda},
\end{eqnarray} 
where $\mathcal{D}(L)$ is the spinor representation corresponding to the pure boost $L$: 
 \begin{eqnarray}
\label{eq:LSR}
\mathcal{D}(L)=\left( \begin{array}{cc} \cosh{\frac{\xi}{2}} & \boldsymbol{\sigma}\cdot \hat{\bf p} \sinh{\frac{\xi}{2}} \\
 \boldsymbol{\sigma}\cdot \hat{\bf p} \sinh{\frac{\xi}{2}} & \cosh{\frac{\xi}{2}} \end{array} \right),
\end{eqnarray}
where $\xi= \tanh^{-1}\beta$ is the rapidity of the particle, $\beta= \sqrt{\boldsymbol{\beta}^2}$, 
and $\hat{\bf p}={\bf p}/\sqrt{{\bf p}^2}$.

The Dirac spin three-vector operator in the particle rest frame is 
known to be 
\begin{eqnarray}
\label{Eq:DSRF}
\tilde{{\bf \mathcal{S}}} = \frac{1}{2}\left( \begin{array}{cc} {\boldsymbol{\sigma}} & 0 \\ 0 & {\boldsymbol{\sigma}} 
\end{array} \right),
\end{eqnarray} 
where ${\boldsymbol{\sigma}}=(\sigma^1, \sigma^2, \sigma^3)$.
To reproduce this Dirac spin three-vector operator in the particle rest frame by using the definition in Eq. (\ref{Eq:SVT}), 
the spin tensor operator in the particle rest frame must be represented by 
\begin{eqnarray}
\mathbb{S}_{\alpha\kappa} |\tilde{p}, \sigma\rangle = \frac{i}{2} [\gamma_\alpha, \gamma_\kappa]|\tilde{p},\sigma\rangle. 
\end{eqnarray}     
A symbol $[,]$ denote the commutation relation. The $|\tilde{p},\sigma\rangle $ is 
the spinor in the particle rest frame which transforms to the spinor in the moving frame by 
$|p,\sigma\rangle= \mathcal{D}(L)|\tilde{p}, \sigma\rangle$. 
Then the spin tensor operator in the moving frame is obtained by the Lorentz transformation as 
\begin{eqnarray}
\mathcal{S}_{\alpha\kappa}|p,\sigma\rangle = \frac{i}{2} L^{\rho}_{\phantom{\mu}\alpha} L^\lambda_{\phantom{\mu}\kappa}
[\gamma_\rho, \gamma_\lambda] |p,\sigma\rangle. 
\end{eqnarray}

By using the relations (\ref{Eq:DVT}) and (\ref{Eq:CRST}), the electric and magnetic dipole moments 
of the Dirac spin in the moving frame can be written explicitly as 
\begin{eqnarray} 
\label{eq:Dipole}
{\bf d}_{\scriptsize{\text{\tiny{D}}}} &=&  
\alpha \left(  {\boldsymbol{\beta}} \times  \tilde{\bf {\mathcal{S}}} \right) - i \alpha\gamma_5 
 \left[ { \tilde{\bf {\mathcal{S}}}}- \frac{\gamma}{\gamma+1}{\boldsymbol{\beta}} ({\boldsymbol{\beta}}\cdot 
{\tilde{\bf {\mathcal{S}}}})\right], \\ \nonumber
{\boldsymbol{\mu}}_{\scriptsize{\text{\tiny{D}}}} &=&  \alpha \left[
{ \tilde{\bf{\mathcal{S}}}}- \frac{\gamma}{\gamma+1}{\boldsymbol{\beta}} ({\boldsymbol{\beta}}\cdot 
{ \tilde{\bf{\mathcal{S}}}})\right] 
+i \alpha \gamma_5 \left[ {\tilde{\bf{\mathcal {S}}}} \times {\boldsymbol{\beta}} \right], 
\end{eqnarray}
where the subscript $\mbox{\scriptsize{D}}$ denote the dipole moments associated with the Dirac spin. 
$\gamma_5$ is defined as 
\begin{eqnarray}
 \gamma_5= i\gamma^0 \gamma^1 \gamma^2 \gamma^3 =\left( \begin{array}{cc}  0 &  \mathcal{I} \\ 
\mathcal{I}& 0 \end{array} \right),
\end{eqnarray}
where $\mathcal{I}$ is a two-by-two identity matrix. 
The differences between these Dirac spin dipole moments and the relativistic spin dipole moments in Eq. (\ref{Eq:RTRDM}) 
are just the $\gamma_5$-proportional terms. 
Note that $\tilde{\boldsymbol{\mu}}= \alpha {\tilde{\bf S}}$ and $\tilde{\bf d}={\bf 0}$. 
In this sense we consider the $\gamma_5$-proportional terms pure quantum mechanical. 
These terms will be shown to be critical for consistency of the covariant SG experiment. 
Because of the $\gamma_5$-proportional term, which is off-diagonal, the representation of the 
Dirac spin operator must be 4-dimensional.

The Dirac spin three-vector operator in the moving frame is obtained as
\begin{eqnarray}
\label{Eq:DSO}
{\bf {\mathcal  S}} = {\gamma}\tilde{\bf {\mathcal S}} - \frac{\gamma^2}{\gamma+1} {\boldsymbol{\beta}} 
({\boldsymbol{\beta}}\cdot 
{\tilde{\bf{\mathcal{S}}}})+i \gamma \gamma_5 \left[ {\tilde{\bf{\mathcal{S}}}} \times {\boldsymbol{\beta}} \right],
\end{eqnarray}  
by using the relation (\ref{Eq:SVT}). 
One can check that the Dirac spin three-vector operators satisfy the commutation relations 
$[\mathcal{S}^i, \mathcal{S}^j]=i \epsilon^{ijk}\mathcal{S}^k$ of the spin group. 
Notice that the relativistic spin three-vector operators in Eq. (\ref{Eq:CRS}) do not satisfy the spin algebra.  
The difference between the Dirac spin operator and the relativistic spin operator is also 
the $\gamma_5$-proportional term in Eq. (\ref{Eq:DSO}). This term is crucial to satisfy the spin algebra. 


\section{SG Experiment}
\label{sec:QSGE}

In this section we will investigate SG experiments for the relativistic spin and the Dirac spin. 
The SG Hamiltonian for the relativistic spin was obtained in the previous section 
by using the relativistic generalization of the classical dipole interactions so that 
we will call the SG Hamiltonian for the relativistic spin the classical SG Hamiltonian. 
On the other hand, the SG Hamiltonian for the Dirac spin will be called the quantum SG Hamiltonian. 
Here it is enough to consider the two reference frames, the particle rest frame and the laboratory frame. 
In the laboratory frame the particle is moving and the SG device is not moving. 
However, the SG device is moving in the particle rest frame. 

In the original SG experiment, a particle with magnetic moment passes through an inhomogeneous magnetic field. 
The inhomogeneous magnetic field is composed of a strong uniform magnetic field and a small non-uniform magnetic field. 
The particle will be deflected into different detectors according to the eigenvalues of the spin 
along the strong uniform magnetic field. 
Therefore, it is enough to consider only the strong uniform magnetic field to determine 
which detector will click and to investigate the spin state in the SG experiment.   

Fig. 1 shows experimental settings in the laboratory frame: 
There is only the magnetic field whose uniform magnetic field is in the positive $y$-direction. 
The particle is moving with the velocity ${\bf v}=v(\cos{\phi}\hat{x} + \sin{\phi}\hat{y})$. 
From now on, we will denote $1$-, $2$-, and $3$-components of the three-vectors as $x$-, $y$-, and $z$-components 
for convenience. 
The angle $\phi$ is the azimuthal angle from the ${x}$-axis in the  ${x}$-${y}$ plane.
The direction of the velocity for the particle in this setting is not usual, 
but will help to investigate the consistency and the covariance of the SG experiment. 

\begin{figure}[htb]
\includegraphics[width=0.4\textwidth]{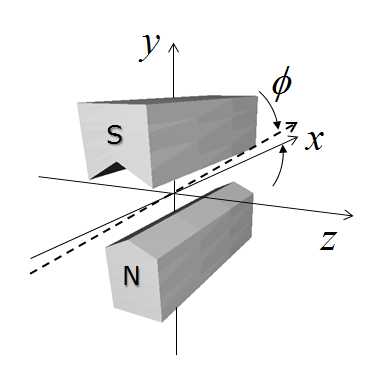}
\caption{This figure is the schematic diagram to show the experimental settings of the SG device. 
The azimuthal angle $\phi$ is the direction of the particle's initial motion.}
\end{figure}

\subsection{SG experiment for relativistic spin}


The classical SG Hamiltonian in the laboratory frame becomes 
\begin{eqnarray}
\label{Eq:USGLAB}
\mathcal{H}_{\mbox{\scriptsize{cSG}}}=- \boldsymbol{\mu} \cdot {\bf B} = - \frac{\alpha}{\gamma} S^y B,
\end{eqnarray} 
where $S^y$ represents the $y$-component of the relativistic spin. 
This shows that the classical SG Hamiltonian and the ${y}$-component of 
the relativistic spin, which is along the magnetic field, have the common eigenstates in the laboratory frame.  

We will obtain the classical SG Hamiltonian in the particle rest frame by using the Lorentz transformation. 
The magnetic field $B  \hat{y}$ in the laboratory frame is transformed into the magnetic field and the electric field 
in the particle rest frame such as
\begin{eqnarray}
\label{Eq:REM}
\tilde{\bf E}&=& -\gamma v \cos{\phi} B {\hat{ \tilde{z}}}, \\ \nonumber
\tilde{\bf B} &=& (1-\gamma)\sin{\phi}\cos{\phi} B { \hat{ \tilde{x}}} + (\sin^2{\phi}+ \gamma \cos^2{\phi})B{ \hat{ \tilde{y}}},
\end{eqnarray}
where the hat symbol $~\hat{ }~$ denotes unit vectors and ${\hat{ \tilde{x}}}$, ${\hat{ \tilde{y}}}$, and 
${\hat{ \tilde{z}}}$ are unit vectors along $\tilde{x}$-, $\tilde{y}$-, and $\tilde{z}$-axis in the particle 
rest frame. 
Since the dipole moments in the particle rest frame are $\tilde{\boldsymbol{\mu}} = \alpha \tilde{\bf S}$ and 
$\tilde{\bf d} = {\bf 0}$, the classical SG Hamiltonian in the particle rest frame becomes 
\begin{eqnarray}
\label{Eq:CSGRF}
\tilde{\mathcal{H}}_{\mbox{\scriptsize{cSG}}} = -\tilde{\boldsymbol{\mu}}\cdot \tilde{\bf B}= - \alpha \tilde{B}^\theta \tilde{S}^{\theta},
\end{eqnarray}
where $\tilde{S}^\theta= \tilde{ S}^x \cos{\theta}+ \tilde{S}^y \sin{\theta}$, 
$\tilde{B}^\theta= \sqrt{\left( \tilde{B}^x\right)^2+ \left( \tilde{B}^y\right)^2}$, and 
\begin{eqnarray}
\label{Eq:TanTheta}
\tan{\theta}= \frac{\sin^2{\phi}+ \gamma \cos^2{\phi}}{(1-\gamma)\sin{\phi}\cos{\phi} }
\end{eqnarray}
The azimuthal angle ${\theta}$ is the angle from the $\tilde{x}$-axis in the  $\tilde{x}$-$\tilde{y}$ plane.

Let us further assume that the magnetic moment of the particle is initially aligned along the direction of the magnetic 
field in the laboratory frame. Because of the relation of ${\boldsymbol{\mu}}= \alpha {\bf S}/\gamma$, 
the direction of the relativistic spin is also initially along the positive $y$-direction in the laboratory frame. 
Notice that the expectation values of the operator can be interpreted as the corresponding classical vectors 
in quantum mechanics. 
Hence the initial state is the eigenstate of the relativistic $S^y$ with eigenvalue $+1/2$ in the laboratory frame. 
The consistency condition for the initial state is that the expectation values of $S^x$ and $S^z$ must be zero. 
For this initial condition the upper detector, which determines the spin is along the positive $y$-direction, will click.

Now let us investigate the consistency and covariance of the SG experiment for the relativistic spin. 
Let us consider the experiment in the particle rest frame. 
In the particle rest frame, the eigenstates of the classical SG Hamiltonian in Eq. (\ref{Eq:CSGRF}) are the eigenstates 
of the $\theta$-component of the relativistic spin three-vector operator in the particle rest frame. 
The eigenstates are 
\begin{eqnarray}
|u_\theta\rangle = \frac{1}{\sqrt{2}} \left( \begin{array}{c} e^{-i \theta /2} \\ e^{i\theta/2} \end{array} \right) 
\mbox{ and } 
|d_\theta\rangle = \frac{1}{\sqrt{2}} \left( \begin{array}{c} e^{-i \theta /2} \\ -e^{i\theta/2} \end{array}\right).
\end{eqnarray}
The expectation values of the components of the relativistic spin in the particle rest frame give the relation \begin{eqnarray}
\label{Eq:MAG}
\frac{\langle u_\theta| \tilde{S}^y |u_\theta\rangle}{\langle u_\theta | \tilde{S}^x |u_\theta \rangle} = 
 = \tan{\theta} , 
\end{eqnarray}
where $\langle u_\theta|$ is the Hermitian conjugate of $|u_\theta\rangle$, i.e., 
$\langle u_\theta|=|u_\theta\rangle^\dagger$. 
 
The covariance of the SG experiment provides the condition of 
$\gamma {\mathcal{H}}_{\mbox{\scriptsize{cSG}}} =\tilde{\mathcal{H}}_{\mbox{\scriptsize{cSG}}}$ 
between the two Hamiltonians in the two reference frames. 
This condition requires that the eigenstates of the two Hamiltonians are the same. 
That is, $|u_\theta\rangle$ and $|d_\theta\rangle$ are also the eigenstates of the classical SG Hamiltonian 
in the laboratory frame. One can easily check that $|u_\theta\rangle$ is the eigenstate of the relativistic spin 
$S^y$ in the laboratory frame with eigenvalue $+1/2$ by using the expression of Eq. (\ref{Eq:CRS}). 
This means that $|u_\theta\rangle$ is the initial state in the laboratory frame.  

Similarly, the expectation value of $S^x$ with the eigenstate $|u_\theta\rangle$ is obtained as 
\begin{eqnarray}
\langle u_\theta | {S}^x |u_\theta \rangle= \frac{(1-\gamma^2)\sin{\phi}\cos{\phi}}{\sqrt{\sin^2{\phi}+\gamma^2 \cos^2{\phi}}}.
\end{eqnarray} 
This cannot be zero in the moving frame where $\gamma\neq 1$. 
This fact is inconsistent to the initial condition in the laboratory frame, which requires that 
the expectation values of $S^x$ and $S^z$ for the initial state of the particle must be zero. 

On the other hand, the consistency condition, that the expectation value of $S^x$ must be zero, gives the relation 
\begin{eqnarray}
\label{Eq:SLT}
\frac{\langle  \tilde{S}^y  \rangle}{\langle  \tilde{S}^x \rangle} = 
\frac{\cos^2{\phi}+ \gamma\sin^2{\phi} }{(\gamma-1) \cos{\phi} \sin{\phi} }=\tan{\xi}.
\end{eqnarray}
Since $\tan{\xi} \neq \tan{\theta}$, the expectation value of the relativistic spin in the particle rest frame must be 
calculated in the state $|u_\xi\rangle$ not the state $|u_\theta\rangle$ in order to satisfy the consistent initial condition. 
This means that the state $|u_\xi\rangle$ must be the eigenstate of the relativistic spin $S^y$ in the laboratory frame. 
The state $|u_\theta \rangle$, however, must be the eigenstate of the classical SG Hamiltonian in the particle 
rest frame. Hence the covariance of the SG experiment, which requires $|u_\theta \rangle=|u_\xi \rangle$, 
is broken for the state $|u_\xi\rangle$.

In summary, the covariance of the SG experiment requires that the eigenstates are the same for both Hamiltonians 
in the laboratory frame and the particle rest frame. 
However, in order to preserve the covariance of the SG experiment, the 
initial condition of the SG experiment in the laboratory frame cannot be satisfied. 
In contrast, to satisfy the consistency requirement of the initial 
condition in the laboratory frame breaks the covariance of the SG experiment. 
This is a paradox.

This paradox between observers in the particle rest frame and the laboratory frame in the SG experiment 
will be resolved in the next subsection by using the Dirac spin.


\subsection{SG experiment for Dirac spin}
\label{sec:DSSG}

Now let us investigate the SG experiment for the Dirac spin. 
The quantum SG Hamiltonian in the laboratory frame is defined by  
\begin{eqnarray}
\label{Eq:SGH}
\mathcal{H}_{\mbox{\scriptsize{qSG}}}= - \frac{\alpha}{2 \gamma}  \mathcal{S}_{\mu\nu}\mathcal{F}^{\mu\nu},
\end{eqnarray} 
similar to the classical SG Hamiltonian. 
Note that the Lorentz transformation of the field tensor, 
\begin{eqnarray}
\mathcal{F}^{\mu\nu}   =  L_{\alpha}^{\phantom{\mu}\mu} 
L_\beta^{\phantom{\mu}\nu} \tilde{\mathcal{F}}^{\alpha\beta}, 
\end{eqnarray}
is defined by the inverse transformation $L^{-1}$. 
This is because the transformation of the Dirac spin tensor operator is defined by the inverse Lorentz transformation 
in Eq. (\ref{Eq:STTR}). 

With the previous experimental setting, the quantum SG Hamiltonian in the laboratory frame 
is written as 
\begin{eqnarray}
\label{Eq:qSG}
\mathcal{H}_{\mbox{\scriptsize{qSG}}} = - \frac{\alpha}{\gamma} \mathcal{S}^y B.
\end{eqnarray}
The initial state of the particle in the laboratory frame is $|p,+y\rangle$, which is the Dirac spinor with eigenvalue +1/2 
for the Dirac spin three-vector operator $\mathcal{S}^y$. 
Then the initial direction of the Dirac spin in the laboratory frame is along the positive $y$-axis. 
The expectation value of the quantum SG Hamiltonian $\mathcal{H}_{\mbox{\scriptsize{qSG}}}$ 
of Eq. (\ref{Eq:qSG}) in the state $|p,+y\rangle$ 
becomes the same as that of the classical SG Hamiltonian $\mathcal{H}_{\mbox{\scriptsize{cSG}}}$ of Eq. 
(\ref{Eq:USGLAB}) in the state $|u_\theta\rangle$. 
Therefore, the detector, which will click in the SG experiment for the Dirac spin, 
must be the same as the upper detector in the SG experiment for the relativistic in the laboratory frame. 

Note that we use the Lorentz invariant normalization $\langle p', \sigma'|\gamma^0|p, \sigma\rangle = 
2p^0 \delta^3({\bf p}' -{\bf p} )\delta_{\sigma' \sigma}$ for the positive-energy Dirac spinors and 
$\langle p', \sigma'|\gamma^0|p, \sigma\rangle = 
- 2p^0 \delta^3({\bf p}' -{\bf p} )\delta_{\sigma' \sigma}$ for the negative-energy Dirac spinors in the 
covariant formalism. 
Here 
\begin{eqnarray} 
\gamma^0= \left( \begin{array}{cc} \mathcal{I} & 0 \\ 0 & -\mathcal{I} \end{array} \right), 
\end{eqnarray}
and $\delta^3({\bf p}' -{\bf p} )$ and $\delta_{\sigma' \sigma}$ are delta function and Kronecker delta, respectively. 
$\langle p, +y|$ is the Hermitian conjugate of $|p, +y\rangle$. 
Notice that the Hermiticity of the Dirac spin three-vector operator in Eq. (\ref{Eq:DSO}) is not preserved in 
the manifestly covariant formalism. 
However, this does not mean the expectation value of the Dirac spin three-vector operator, which is defined by 
$\langle p,\sigma |\gamma^0 \mathcal{S}^i |p,\sigma\rangle$ for the positive energy state and 
by $-\langle p,\sigma |\gamma^0 \mathcal{S}^i |p,\sigma\rangle$ for the negative energy state, is not real. 
The real-valuedness of the expectation value is guaranteed by the fact that $\gamma^0 \mathcal{S}^i$ is Hermitian. 
We have shown in the previous work \cite{Choi1} that the expectation value of the Dirac spin three-vector operator 
in the covariant formalism is the same as the expectation value of the Foldy-Woutheysen mean spin operator, 
which is Hermitian, in the canonical formalism. 
In the canonical formalism the Hermiticity of the spin three-vector operator 
is preserved and the expectation value is defined as the normal norm of the Hilbert space. Hence one can use 
the canonical formulation if one want to follow the familiar formalism, the same as the non-relativistic quantum 
mechanics. 
In this paper, however, we are interested in the covariance of the SG experiment, so the covariant formalism using 
the Dirac spinors are most convenient. 

The expectation value of the $x$-component of the Dirac spin will automatically zero for the Dirac spinor $|p,+y\rangle$, because 
${\mathcal{ S}^x} |p,\sigma\rangle= \mathcal{D}(L) \tilde{\mathcal{S}}|\tilde{p},\sigma\rangle$. 
Hence, the consistency requirement of the initial condition for the SG experiment in the laboratory frame is 
satisfied for the Dirac spinor $|p,+y\rangle$.

The condition of the covariance of the SG experiment for the Dirac spin requires that 
the eigenstates of the two quantum SG Hamiltonians, $\tilde{\mathcal{H}}_{\mbox{\scriptsize{qSG}}}$ and 
${\mathcal{H}}_{\mbox{\scriptsize{qSG}}}$, must be the same. 
Hence the Dirac spinor $|p,+y\rangle$ must also be the eigenstate of the quantum SG Hamiltonian in the particle rest frame. 
The quantum SG Hamiltonian in the particle rest frame is obtained by the Lorentz transformation as 
\begin{eqnarray}
\label{Eq:QSGRF}
\tilde{\mathcal{H}}_{\mbox{\scriptsize{qSG}}}&=& - \tilde{\boldsymbol{\mu}}_{\scriptsize{\text{\tiny{D}}}}\cdot 
\tilde{\bf B} - \tilde{{\bf d}}_{\scriptsize{\text{\tiny{D}}}} \cdot \tilde{\bf E} \\ \nonumber
&=& -\alpha \tilde{\bf {\mathcal {S}}} \cdot \tilde{\bf B} +i \alpha 
\gamma_5 \tilde{\bf {\mathcal{S}}}\cdot \tilde{\bf E},
\end{eqnarray}
where $\tilde{\bf E}$ and $\tilde{\bf B}$ are in Eq. (\ref{Eq:REM}). The $+$ sign for the electric dipole interaction term 
stems from the fact that we have used the inverse Lorentz transformation. 

The electric dipole moment of the Dirac spin in the particle rest frame has nonzero $\gamma_5$-proportional term. 
Hence the quantum SG Hamiltonian in the particle rest frame 
has the nonzero electric dipole interaction $i \alpha \gamma_5 \tilde{\bf {\mathcal{S}}}\cdot \tilde{\bf E}$ 
even in the particle rest frame. 
This is the critical difference from the classical SG Hamiltonian in Eq. (\ref{Eq:CSGRF}). 
This implies that the direction of the Dirac spin in the particle rest frame cannot be determined solely 
by the magnetic field at the particle rest frame. 
Note that the expectation value of the electric dipole interaction becomes
\begin{eqnarray}
\langle p, +y|\gamma^0(i \alpha \gamma^5 \tilde{S}^z E^z)|p,+y\rangle =(\gamma^2-1)\cos^2{\phi} B,
\end{eqnarray}
where the positive-energy state is considered. Note that the expectation value for the negative-energy state 
only changes the sign. 

Let us consider the direction of the Dirac spin in the particle rest frame for the positive-energy state. 
The expectation values of the Dirac spin in the particle rest frame is obtained as  
\begin{eqnarray}
\label{Eq:RSEX}
\langle p, +y|\gamma^0\tilde{\mathcal{S}}^x|p, +y\rangle &=& (1-\gamma) \cos{\phi}\sin{\phi}, \\ \nonumber
\langle p, +y|\gamma^0\tilde{\mathcal{S}}^y|p, +y\rangle &=& \sin^2{\phi}+\gamma \cos^2{\phi}, \\ \nonumber
\langle p, +y|\gamma^0 \tilde{\mathcal{S}}^z|p, +y\rangle &=& 0.
\end{eqnarray}
The above relation shows that the direction of the Dirac spin in the particle rest frame, which is determined by 
$\langle p, +y|\tilde{\mathcal{S}}^y|p, +y\rangle / \langle p, +y|\tilde{\mathcal{S}}^x|p, +y\rangle$, 
is along $\theta$-direction. 
This means that the both directions determined by the relativistic spin and the Dirac spin in the particle rest frame 
are the same in the SG experiment. 

The expectation value of the $x$-component of the Dirac spin in the laboratory frame, however, 
is zero in the eigenstate of the quantum SG Hamiltonian in the particle rest frame, which is $|p,+y\rangle$. 
Therefore, the Dirac spin resolves the paradox. That is, the Dirac spin satisfies the covariance of the 
SG experiment which requires the same eigenstates for the quantum SG Hamiltonians in the 
particle rest frame and the laboratory frame. And the Dirac spin also satisfies the consistency relation 
which is required by the initial condition.  

The investigation of the covariant SG experiment for the Dirac spin shows that the Dirac spin 
is clearly determined in the frame, where only magnetic field exists, by the SG experiment. 
Notice that the interpretation of the SG experiment for the Dirac spin in the laboratory frame is the same 
as that in the original SG experiment for the non-relativistic spin, because the form of the Hamiltonian is the same. 
This means that the Dirac spin can be determined independent on the momentum of the particle 
by the standard SG experiment, where there is only magnetic field. 
This implies that the density matrix for the Dirac spin can be well-defined by integrating out the momentum 
degrees of freedom contrary to previous works.

\section{Conclusion}

We have considered the covariant SG experiments for the relativistic spin and the Dirac spin. 
The relativistic spin was obtained by the classical spin dipole tensor which gives 
the covariant extension of the spin dipole interactions under the Lorentz transformation. 
The relativistic spin is a natural relativistic extension from the non-relativistic spin in this sense. 
Hence the relativistic spin seems to be the best candidate for the spin of a massive relativistic particle 
in an arbitrary frame.  
However, we have shown that the relativistic spin has crucial problems to be a proper spin operator. 
Most of all, the relativistic spin cannot satisfy the spin algebra. 
Moreover, the covariant SG experiment for the relativistic spin is shown to provide a paradox between observers 
in the particle rest frame and the laboratory frame. 

This is because the spin for a massive relativistic particle is not only relativistic but also quantum mechanical. 
We have shown that the Dirac spin, which is covariantly defined in the Dirac theory, 
not only satisfies the spin algebra but also resolve the paradox in the SG experiment for the relativistic spin. 
Moreover, the directions of the two spins, which are determined by the expectation values of the relativistic spin 
and the Dirac spin in the same settings of the SG experiment, are the same. 
The additional term of the Dirac spin is the off-diagonal term which can be interpreted as a pure quantum mechanical 
contribution. The expectation value of the off-diagonal term is higher order than the diagonal term, because 
it mixes the large and the small components of the Dirac spinor. 
Our results suggest that the Dirac spin operator is the proper spin operator in nature.

The Dirac spin in the laboratory frame can be determined by the magnetic field in the 
standard SG experiment independent on 
the momentum of the particle. This shows that the spin reduced density matrix 
for a massive Dirac particle can be clearly defined by integrating out momentum degrees of freedom.   


\section*{Acknowledgements}

 This work was supported by National Research Foundation of Korea Grant funded by the Korean
 Government(2012-0003786). Y. D. Han was supported by Woosuk University. 
 We gratefully acknowledge KIAS members for helpful discussions.


\end{document}